\documentclass[11pt,a4paper]{article}

\usepackage[T1]{fontenc}
\usepackage[utf8]{inputenc}
\usepackage{microtype}
\usepackage{helvet}  
\usepackage{amsmath,amssymb}
\usepackage{graphicx}
\usepackage{xcolor}
\usepackage[top=2.5cm,bottom=2.5cm,left=2.5cm,right=2.5cm]{geometry}
\usepackage{soul}

\usepackage[numbers]{natbib}
\definecolor{natureblue}{RGB}{0,75,135}
\definecolor{naturedarkgray}{RGB}{51,51,51}

\usepackage{titlesec}
\titleformat{\section}
  {\normalfont\large\bfseries\color{naturedarkgray}}
  {}{0em}{}
\titlespacing*{\section}{0pt}{1.5em}{0.5em}

\usepackage{hyperref}
\hypersetup{
    colorlinks=true,
    linkcolor=natureblue,
    filecolor=natureblue,
    urlcolor=natureblue,
    citecolor=natureblue,
    pdftitle={What Understanding Means in AI-Laden Astronomy},
    pdfauthor={Yuan-Sen Ting, André Curtis-Trudel, Siyu Yao},
}

\makeatletter
\renewcommand{\maketitle}{%
  \begin{flushleft}
    {\fontsize{20}{24}\selectfont\bfseries\color{naturedarkgray} What Understanding Means in AI-Laden Astronomy\par}
    \vspace{1em}
    {\large Yuan-Sen Ting\textsuperscript{1,2,3}{\Large *}, André Curtis-Trudel\textsuperscript{4,5}, Siyu Yao\textsuperscript{6}\par}
    \vspace{0.8em}
    {\small\color{gray}
    \textsuperscript{1}Department of Astronomy, The Ohio State University, Columbus, OH, USA.\\
    \textsuperscript{2}Center for Cosmology and AstroParticle Physics (CCAPP), The Ohio State University, Columbus, OH, USA.\\
    \textsuperscript{3}Max-Planck-Institut f\"ur Astronomie, Heidelberg, Germany.\\
    \textsuperscript{4}Department of Philosophy, University of Cincinnati, Cincinnati, OH, USA.\\
    \textsuperscript{5}UC Center for Humanities and Technology, University of Cincinnati, Cincinnati, OH, USA.\\
    \textsuperscript{6}Department of Philosophy, School of Humanities, Shanghai Jiao Tong University, Shanghai, China.\par}
    \vspace{0.3em}
    {\small\color{gray} {\Large *}e-mail: ting.74@osu.edu; andre.curtis-trudel@uc.edu; yaosy@sjtu.edu.cn\par}
  \end{flushleft}
  \vspace{1em}
}
\makeatother

\renewenvironment{quotation}
  {\list{}{\leftmargin=2em\rightmargin=2em}\item\relax\itshape}
  {\endlist}

\begin{document}

\maketitle

The British philosopher Mary Midgley once remarked that philosophy is like plumbing: you don't notice it until things start to smell funny. And astronomers are beginning to sniff the air. As AI transforms astronomical research, scientists are confronting foundational philosophical questions about the nature of discovery and understanding—questions that call for collaboration with philosophers of science.

The hype surrounding AI in science is not difficult to find. What is harder to locate is sustained, rigorous reflection on what AI actually changes about how we generate knowledge. In the rush to deploy ever more powerful models, the scientific community often treats AI integration as a purely engineering challenge—a matter of optimizing workflows and accelerating outputs. Yet beneath these practical concerns lie deeper questions that cannot be resolved by better algorithms alone. What does it mean to understand a phenomenon? What amounts to genuine discovery, and how does AI facilitate it beyond producing mere predictions? How should the community evaluate scientific contributions in an era where much of the labour can be automated?

Philosophers and historians of science have long grappled with these questions, yet they are often left out of mainstream discussions about AI in research, as if the transformation were purely a matter for engineers to optimize and scientists to exploit. This is a missed opportunity. Philosophy offers astronomy three things: conceptual engineering (What do we mean by `understanding'?), critical examination of assumptions (Why do we believe data from the real world is privileged?), and frameworks for abstraction (How do AI's roles differ across subfields?). These are not luxuries. They are essential tools for navigating a transformation already underway.

To foster this dialogue, we recently convened an interdisciplinary workshop---``Philosophy Sees the Algorithm''---assembling astronomers, philosophers, computer scientists, and social scientists. The workshop, organized by Siyu Yao, Yuan-Sen Ting, and Andr\'e Curtis-Trudel, was held on 11--12 December 2025 at The Ohio State University (see \url{https://tingyuansen.github.io/Philosophy_in_AI_Based_Science/}). The idea of understanding emerged as the centre of various questions and dialogues: the understanding of science, in science, and with science. What follows draws on those conversations, but the questions belong to everyone navigating this transformation.

\section*{The productivity revolution and its discontents}

Productivity gains from AI are real and accelerating. Complex statistical analyses that once required months can now be completed in days. This acceleration shows no sign of stopping, and its reach is growing wider. The implications extend beyond efficiency. When coding, mathematical derivation, and literature synthesis can be substantially automated, we must ask: what remains distinctively human about astronomical research?

In addressing this question, it is crucial that we accurately characterize astronomy's goals and methods. Modern astronomy is observation-driven. The discoveries that advance our field are typically throttled principally by our ability to collect observations—by instrumentation, survey design, and the gradual expansion of our observational frontier. The James Webb Space Telescope and gravitational wave detector LIGO exemplify this reality: breakthroughs emerge from finding mismatches between existing theory and newly collected data with advanced observational capabilities. Although AI can support these endeavours, it is far from clear that AI on its own can fully replace them.

This feature of modern astronomy helps explain the feeling that AI tools are out of step with modern astronomy's actual needs and concerns. Papers celebrating AI's ability to `rediscover Newtonian laws' from data are frequently cited in machine learning literature. Yet these contributions appear to rest on an outdated conception of astronomical research, one resembling something like 18th century natural philosophy in which the goal is to derive fundamental equations through symbolic manipulation. It is unsurprising that such contributions have made little impact on actual astronomical research.

In thinking about how AI might contribute to a more realistic vision of astronomy, we should also be clear-eyed about AI's capabilities and limitations and alert to new risks it might pose. Machine learning excels at sifting through vast archives for missed patterns, and such applications have already yielded scientific insights—much as astronomers have continued to extract novel discoveries from the Hubble Space Telescope's decades-old data. At the same time, delegating crucial elements of astronomical research to AI might diminish astronomers' capacity for observational innovation, including designing and evaluating proposed observational methods.

Recent philosophy of astrophysics has emphasized these `non-lawful' aspects of astronomical knowledge: developing observations and measurements, determining the existence of entities, astronomy as a historical science describing particular evolutionary scenarios, and astronomy as natural history interested in comprehending the nature of individual objects and their population \citep{MillsBoyd2023}. These dimensions of astronomical practice are poorly captured by the equation-derivation model—and we contend that they are precisely the dimensions where human judgment remains essential.

\section*{Understanding beyond equations}

Even setting aside the data-collection bottleneck, recent philosophical work emphasizes that there's more to scientific understanding than deriving equations \citep{Schickore2025}. Consider galaxy evolution. We possess excellent knowledge of the fundamental forces governing galactic dynamics—gravity, hydrodynamics, radiative processes. We can run sophisticated simulations that reproduce observed phenomena with remarkable fidelity. Yet our `understanding' of galaxy evolution is not simply the recovery of Newton's equations. It encompasses grasping the emergent behaviours of complex systems, the particular evolutionary pathways that give rise to observed structures, and the relationships between scales and environments.

This observation connects to longstanding philosophical debates about the nature of scientific understanding. Philosophers distinguish between different epistemic goods that science pursues: prediction, explanation, manipulation, and understanding \citep{Kuhn1977}. These are related but not identical. A model that predicts accurately may not provide understanding; a theoretical framework that explains phenomena may not enable prediction; an idealization may fail to predict or explain real phenomena, but it can help one grasp their underlying salient causal relations. The question of what constitutes understanding—and whether AI can achieve or facilitate it—is not merely academic. It has practical implications for how we evaluate research, train students, and allocate resources.

While various ways of thinking about scientific understanding are possible, one particularly useful conception of understanding emphasizes several interconnected capacities: characterizing the features of a system, communicating those characteristics so that others can mentally reconstruct them, and (where applicable) manipulating or controlling the system \citep{Potochnik2017}. This framework highlights an often-overlooked dimension of scientific knowledge emphasized by philosophers and sociologists of science: its essentially communicative and social character \citep{Longino1990}. On this way of thinking, understanding is a matter of making the world intelligible to communities of inquirers.

\section*{The narratives of science}

If understanding is essentially communicative, then narrative matters. This brings us to what may seem an unlikely topic for an astronomy perspective: narrative. Telling good stories is integral to science, because a compelling scientific narrative does more than report existing findings; it enables discovery by connecting and making sense of evidence from scattered sources, situates an outcome within a web of context, identifies what is surprising or important, draws connections across domains, and guides future inquiry \citep{Fletcher2023}. The social character of scientific knowledge means that understanding must be transmissible; it must be capable of being reconstructed in other minds. This is fundamentally an expressive achievement.

Consider a thought experiment from literary studies: the power of the six-word story. The famous example attributed to Hemingway—`For sale: baby shoes, never worn'—conveys vastly more than its six words literally contain. The emotional weight, the implied narrative, the contextual knowledge readers bring to bear are essential to the story's meaning. Scientific understanding in complex domains shares something of this character. The `knowledge' encoded in a successful model of galaxy formation is not fully captured by its equations or even its predictions; it includes the tacit understanding of which features matter, why they matter, and how they connect to the broader enterprise of astronomy.

Current large language models demonstrate impressive capabilities in certain domains—generating plausible text, summarizing literature, and assisting with routine tasks. But they struggle with precisely this kind of contextual, narrative reasoning. They can produce outputs that superficially resemble scientific prose, but it is considerably less clear whether they can grasp what makes a scientific story compelling or what connections would genuinely advance understanding. Of course, it remains to be seen whether this limitation is merely technical; whether next-token predictors can somehow come to `grok' these elements of scientific thought or whether it reflects a more fundamental limitation on their capabilities.

\section*{Problem-solving versus problem-finding, and the limits of AI}

A useful distinction here comes from Herbert Simon's work on the structure of problems. Simon distinguished between well-structured and ill-structured problems \citep{simon1973structure}. Well-structured problems have clear goals, well-defined methods for solution, and unambiguous success criteria. Examples include solving a system of equations or optimizing a function. Ill-structured problems lack this structure; they require formulating the problem itself before any solution can be attempted. Much of what makes research difficult is not solving problems but figuring out which problems are worth solving and how to frame them.

This distinction illuminates both AI's current contributions and its limitations. Large language models are increasingly adept at well-structured tasks: given sufficient training data, they can reliably complete mathematical derivations, generate code, produce text, or solve complex subject competition problems, which follow or go beyond conventional patterns \citep{Hubert2025,Novikov2025,Pinheiro2025}. These are genuine achievements, and because so much daily research labour involves such tasks, AI assistance can dramatically accelerate progress.

Yet scientific breakthroughs rarely emerge from solving well-structured problems more efficiently. They emerge from recognizing a puzzling phenomenon, that existing frameworks fail to accommodate some phenomenon, and that new tools are needed. Often, it is appropriate to think of these cases as involving problem finding rather than problem solving. And the task of problem finding is frequently ill-structured in Simon's sense. It is therefore unsurprising that experienced researchers report that AI often falls short at this task. The deeper work of identifying promising research directions, recognizing when data genuinely challenges existing frameworks, and constructing narratives that advance community understanding seems to require something beyond pattern recognition within established problem structures.

Whether AI might eventually develop such capacities remains under speculation. For the time being, questions about division of labour remain central. On the one hand, we may wish to hand off to AI the problem-solving work that constitutes much of research's daily work, while the problem finding that gives research its direction remains distinctively human. This approach might accelerate problem solving, but might also cause scientists to lose the expertise required for problem finding. On the other hand, we can imagine a more limited role for AI, one which assigns only the most rote, routine tasks, leaving much else for humans. Naturally, other approaches are possible, although our instinct is that leaving humans entirely out of the loop is premature.

\section*{Pursuitworthiness and scientific acceleration}

In thinking about how AI can accelerate the solution of well-defined problems, it is useful to distinguish three aspects of scientific work \citep{Laudan1978}: generation (How do scientists produce ideas?), pursuit (How do they decide which ideas to work on?), and acceptance (How do ideas become established knowledge?). Much discussion of AI focuses on generation and acceptance. But the middle stage, pursuit, may be where AI's impact is most profound and least examined.

Given infinite time, we might pursue all questions. But human time is finite, and we must be selective. This selection is not arbitrary; it is guided by criteria that the scientific community has developed, often tacitly. Philosophers of science have articulated some of these. Is an idea minimally plausible? Is it interesting? Does it fit the available evidence? Would it unify or explain? These criteria help researchers judge what makes it reasonable to work on one idea over another.

However, in addition to these, economic considerations play an important role in determining whether to pursue an idea. The American scientist and philosopher C. S. Peirce puts this vividly:

\begin{quotation}
Proposals for hypotheses inundate us in an overwhelming flood \ldots\ the process of verification to which each one must be subjected before it can count as at all an item, even of likely knowledge, is so very costly in time, energy, and money \ldots\ that Economy would override every other consideration even if there were any other serious considerations. \citep{Peirce1958}
\end{quotation}

Here may be one of AI's most profound effects. As AI tools abolish much of the mechanistic labour, drastically shortening project timelines, the landscape of what is feasible shifts. Topics that once seemed prohibitively labour-intensive become tractable. This expansion of possibility is genuinely exciting. But it also raises concerns. If the criteria for pursuitworthiness remain tacit and unarticulated, they may shift without conscious deliberation—pulled by the gravitational force of what AI makes easy rather than what is genuinely important. Remaining cognizant of this incentive is an ongoing concern for any scientific community adopting AI \citep{Messeri2024}.

\section*{Peer review and the filtering problem}

In the limit where AI can perform low-hanging mechanistic tasks and produce passable papers at scale, the risk of flooding the literature becomes real. If publication counts and citation metrics have traditionally served as proxies for contribution, what happens when those proxies become decoupled from genuine understanding? The sheer volume of AI-assisted outputs could overwhelm our capacity to identify breakthrough science.

This is where peer review and grant evaluation become critical. Recent research has found that proposals and papers with extensive AI-generated content—where ideas themselves are AI-produced or where writing relies heavily on AI generation—may be less favourably evaluated in certain dimensions than conventional approaches \citep{Eger2025}. While such findings admit of multiple interpretations, they suggest that human reviewers remain attuned to something beyond surface productivity. They are looking for the sparks—the signs of genuine insight, the evidence that a researcher understands not just how to generate results but why those results matter.

Human-based reviewers, however imperfect, may be our best option for maintaining the health of the scientific enterprise. Precisely because context matters, because narrative matters, because understanding in complex systems cannot be reduced to outputs alone, the judgment of experienced researchers remains essential. The question is whether our institutions can adapt quickly enough to preserve this function as the volume of AI-assisted work accelerates.

\section*{Towards pragmatic understanding}

How, then, should the astronomy community navigate this transition? The concept of `pragmatic understanding' offers a useful framework. On this view, understanding is not an all-or-nothing matter of grasping fundamental truths, but a practical achievement that admits of degrees and depends on context and purpose. To possess pragmatic understanding in a domain means having the skills to wield one's knowledge, tools, and evaluation standards towards the reliable achievement of scientific aims \citep{chang2022realism}.

Consider the historical parallel of computer simulations. Two decades ago, researchers who relied heavily on simulations were sometimes dismissed as not doing `real' theory—genuine understanding required analytic solutions. This attitude has shifted. The community now recognizes simulations as legitimate components of the knowledge system, even as we acknowledge their limitations and the need for ongoing validation against observations. Simulations turn out to enhance astronomers' pragmatic understanding of their models and theories by bridging them to complex phenomena for validation and application.

AI may follow a similar trajectory. Large models that extract latent representations from astronomical data, even if their internal workings remain opaque, may come to function as tools for pragmatic understanding—not replacing human insight but extending it, providing new ways of characterizing complex phenomena that complement traditional approaches. To do this, the astronomy community also needs to establish an understanding of AI tools by articulating systematic lessons about how AI can be properly used for different kinds of questions, target systems and purposes. The key is developing appropriate norms for validation, integration, and communication.

What counts as AI fluency or literacy? What data infrastructures—curation, metadata, provenance tracking, reproducibility pipelines—are essential for meaningful AI use? When is AI an exploratory tool versus part of the evidential chain? Are AI tools opaque in novel or unanticipated ways? How do novel forms of opacity complicate peer review? These questions require collaboration between scientists and scholars who study science—philosophers, historians, and social scientists who can help articulate what we value, identify underlying assumptions, and develop evaluation frameworks that keep pace with change.

\section*{A new kind of interlocutor}

AI may represent the first `other intelligence' we can meaningfully communicate with—an entity that talks in ways we may potentially understand, yet whose cognition differs fundamentally from our own. This does not mean AI intelligence equals human intelligence, let alone that AI knowledge equals human knowledge. But it does mean we face something genuinely novel.

We are not simply adopting better instruments. We are learning to collaborate with systems that process language, generate outputs, and respond to queries in ways that require interpretation rather than mere reading. The pragmatic understanding we developed for simulations—accepting tools that extend cognition while remaining clear about their proper use and limitations—must now extend to a new kind of interlocutor.

This is why the philosophical questions are not optional. What does it mean to understand a phenomenon when the tool that helped you get there cannot itself understand? What does authorship mean when contributions blur between human and machine? What does progress mean when traditional academic metrics no longer reliably track the goods we care about?

Philosophy sees the algorithm. And in attending to the plumbing, we may discover what astronomy has always tacitly known: that understanding the Universe is a distinctly human project—even when, especially when, we have non-human collaborators in the endeavour.

\vspace{1.5em}
\noindent\rule{\textwidth}{0.5pt}

\section*{Acknowledgements}
We thank the speakers and panelists at the ``Philosophy Sees the Algorithm'' workshop for their contributions to the discussions that informed this perspective: M.\ Andrews, T.\ Berger-Wolf, B.\ Carstens, A.\ Fletcher, E.\ Fosler-Lussier, J.\ Greenberg, M.\ Hosseinioun, K.\ Kish Bar-On, H.\ Meskhidze, P.\ Osmer, J.\ Phelan, B.\ Santos Genta, and D.\ Weinberg. The workshop webpage is at: \url{https://tingyuansen.github.io/Philosophy_in_AI_Based_Science/}. This work was supported by the Alfred P. Sloan Foundation (award G-2025-25315), the Center for Cosmology and AstroParticle Physics at The Ohio State University, the National Endowment for the Humanities (award RAI-306945-26), and the UC Center for Humanities and Technology.

\section*{Competing Interests}

The authors declare no competing interests. 

\vspace{2em}
\noindent\rule{\textwidth}{0.5pt}

\bibliography{references}

\end{document}